# Review on the stabilization of non linear systems achieved by output feedback control technique

Asifa Yousaf

*Control Science and Engineering, School of Automation, Central South University, Changsha 10533, China*

**I. Abstract—**Stability and control of a non-linear system represent an important system configuration that frequently arises in practical engineering. Stability covers a vast range of systems that do not obey the superposition principle and applies to more real-world systems because all real control systems are non-linear. For efficient stabilization of these systems, a great number of researches have been proposed. This paper surveys some well-known facts as well as some recent developments and different strategies for the topic of stabilization of non-linear systems by output feedback control techniques.

## II. Introduction

Non-Linear system's Control and Stability analysis is one of the most important topics for the past many years and has attracted lots of researchers as discussed in survey of [1, 2] compared to linear systems. It offers a wide range of applications and plays a significant role in real-life controllability problems such as stabilization, noise rejection, and tracking. It is a difficult task to obtain the desired response from non-linear systems and extensive research has been conducted on the stability of these systems such as lyapunov direct method described in, e.g., [3] and stability on equilibrium point of the system in, e.g., [4] and also different control techniques listed in the survey of [5]. Non-linear systems contain non-linearity between their input and output characteristics. Output feedback configuration, also known as closed loop systems is the most important control technique used for non-linear systems. In a feedback control system corrections are made by the feedback of the current output. Feedback could be a positive or negative feedback system where output values are either added or subtracted from the reference value. The controlling technique is categorized into non-linear and linear control systems. This paper surveys the several past and recent theoretical results on the stabilization of nonlinear systems by output feedback control.

*\*The author is with School of Automation, Department of Control Science and Engineering \* Email: 214618007@csu.edu.cn*

## III. Literature Review

For better stabilization, the output of these systems keeps updating its input for modified system response. The impact of feedback on the output response depends on the design and formulation techniques of non-linear control problems and approach methods that are found in classical control theory as described in [6, 7] where feedback stabilization design was proposed for non-linear systems and a controller was designed based on the choice of proper Lyapunov functions to ensure the asymptotic stability of the closed loop system. The control of linear dynamic systems using adaptive fuzzy controller has been analyzed in [8] that operated in the direct adaptation mode with output feedback and with a linear or non-linear model, which determined the desired response of closed loop system.

The problem of output feedback control is more challenging than that of the state feedback control because the system states are incompletely measurable. In the last few years, many output feedback control techniques has been developed for the stabilization, regulation and synchronization of non-linear systems. The problem of output synchronization of non-linear multi agent systems has been solved by the measurement output feedback control technique in [9, 10]. Furthermore, robust output synchronization of non-linear heterogeneous multi agent networks has also been achieved by output feedback control in, e.g., [11-14]. To solve the global robust servomechanism problem for uncertain non-linear systems [15] proposed a solution by output feedback control. General framework was proposed in [16, 17] to solve the semi global and global robust output regulation problem for non-linear systems by output feedback. In particular, the output regulation problem of singular non-linear system has been addressed in [18] by normal output feedback control.

Also memory less feedback controllers has been widely used for the stabilization of several non-linear systems with time delay effects. The problem of stability and stabilization for neutral delay-differential systems with infinite delay has been investigated in [19]. By using Lyapunov-Krasovskii approach, the delay independent stability conditions were obtained for neutral systems which were then represented in terms of linear matrix inequality (LMI) and solved using the Matlab's LMI toolbox. The stabilization of system was achieved by memory-less feedback controller that was also expressed in terms of LMI and solved using Matlab's LMI toolbox.

Moreover, the global state regulation can be achieved in a class of non-linear parametrized time delay systems by memory-less output feedback stabilization controllers with dynamic gains as designed in [20]. Under a parametrized high-order output growth condition with a time delay, the problem of global adaptive regulation via output feedback has been solved in this study. The growth conditions in non-linear perturbations were more general than those found in previous studies. The dynamic linear and memory-less output feedback controller was designed by the Lyapunov-krasovskii function that could achieve stabilization in the presence of disturbances and time delay.

Output feedback control technique along with power integrator and observer design has been analyzed by many researchers to achieve stabilization of certain non-linear systems. Such as, for nonlinear systems with a time-varying power and unknown continuous output function, by constructing a new nonlinear reduced-order observer together with adding a power integrator method, a new function to determine the maximal open sector of output function is given in [21]. This study proposed a technique to design an output feedback controller for the non-linear system having unknown continuous output functions which was not studied before. The designed output feedback controller ensured that the equilibrium point of the close loop system is globally asymptotic stable when the output function belongs to any closed sector included in maximal open sector $\Omega$, which is determined by reduced order non-linear observer containing power integrator.

Similarly, Output feedback control techniques containing power integrator can be applied to second order non-linear systems having high and low order uncertainties to achieve global stabilization as in, e.g., [22]. The feedback controller designed with a power integrator and a gain function can overcome the increase of non-linearities in high and low-order systems. The assumptions presented in this article were more general compared to [23, 24] because the results obtained in previous studies for output feedback stabilization were small and had complexities while designing an observer for these systems. The global asymptotic stability of the feedback system has been achieved by the new Lyapunov function. This note provided a better solution to design the feedback control law as compared to the reduced-order design approach.

The issue which the researches concentrate a lot on it is to deal with unknown growth rate and disturbances which are found widely in practical non linear systems. Instead of a time-varying approach to handle the unknown growth rate, an adaptive output feedback stabilizer for a class of stochastic feed-forward systems possessing an unknown growth rate was constructed in [25] by using the dynamic gain method, By introducing an output feedback stabilizer with an unknown growth rate, this research overcome the limitations of previous literature which was restricted to conditions on non-linear growth that means the growth rate is assumed to be a known constant. Unmeasured system states were estimated by dynamic high gain. Adaptive output feedback controller ensured that all signals of the closed loop system were bounded and the system states converged to zero.

The general framework was also introduced in [26] for a class of high-order nonlinear systems with additive input disturbances and unknown growth rates by global output-feedback stabilization. The additive input disturbances were assumed to be bounded by an unknown positive constant, and hence may not be periodic or with explicit expression. Input disturbances were compensated by the time-varying method developed in [27]. A time-varying output feedback control was designed to deal with unknowns. Signals of the closed-loop system were bounded by selecting suitable parameters in the observer, and system and observer's state converged to zero.

It can be inferred that the output feedback controllers have proved to be an efficient solution to solve the problem of stabilization for non-linear systems containing uncertainties. For example, for a class of non-linear systems containing uncertain control coefficients and unmeasured states dependent growth, an output feedback controller was designed in [28] to achieve global asymptotic stabilization. This note further investigated global asymptotic stabilizing control by output feedback studied before in [29] with updated state transformation and observer design. According to this note, the global asymptotic stabilization can be obtained by choosing the design parameters properly. A simple method was presented for the controller design which reduced the dimension of the closed loop system from 3n to 2n. Stabilization of closed loop systems was achieved by selecting appropriate design parameters.

Furthermore, the output feedback control technique can also be applied to feedforward non-linear systems with unknown control coefficients for semi global finite-time stabilization as in, e.g., [30]. To achieve semi-global stabilization, weaker growth conditions can be imposed on the non-linearities. By supposing non-linear terms to be bounded by nonhomogeneous-type function in the unmeasured states with nonconstant growth rate, a new controller has been designed for general feed-forward systems. In addition, a homogeneous observer with an improved scaling gain was constructed with an appropriate scaling change that addressed the contradiction between the small scaling gain employed for the feed forward system and the large scaling gain used for semi global stabilization. A new coordinate change was introduced to deal with upper triangular growth conditions by scaling gain.

Similarly, the global output feedback stabilization has been also solved for a class of uncertain non-linear systems with unmeasured states dependent, output-dependent polynomial growth and uncertainties in the system dynamics and output equations in [31]. The assumptions were more general compared to the related work in [32-34]. A less dimensional closed loop controller has been designed based on a high gain observer consisting of two gain components that efficiently minimized uncertainty in output-dependent polynomial growth and output equations and offset the unknown growth rate.

Output feedback stabilization for a class of non-linear systems with a higher relative degree can be achieved by using dynamic gain as in, e.g., [35] where global stabilization was considered by utilizing a high gain approach with the following assumption for non-linear terms.

$$|\phi_i(x, u, t)| \leq c(y)(|x_1| + |x_2| + \cdots + |x_i|), 1 \leq i \leq n \quad (1)$$

Where $\phi_1, \phi_2, \cdots, \phi_n$ are the nonlinear terms, $x_1, x_2, \cdots, x_n$ are the states, u is input, and c is a known positive constant. The output feedback controller has been designed by using high gain, to achieve stability. Global stabilization of closed loop was obtained by the dynamic high gain observer and established a control law through the observer. Simulation results for a second order non-linear system verify the effectiveness of proposed control designs.

Non-linear feedback controllers are frequently implemented using linear actuators, which have faster dynamics than nonlinear dynamics of plant and controller. The question is, how faster should the dynamics of the linear actuator be to ensure stability of the given non-linear feedback control system? This note [36] examines the underlying stability difficulties and provides quantitative measurements for linear actuators used in non-linear applications. The stability of the closed loop system has been analyzed by two methods, the Small gain theorem developed in [37] and [38] and the Singular perturbation method [39], and provided sufficient conditions for the actuator to get asymptotic stability of the systems. High gain feedback ε implies the fast response of the actuator feedback loop that must be greater than zero for a system to be asymptotically stable. The results of this note provide guidance to selection and design of linear actuators to be implemented in the non-linear feedback systems, which are initially designed in absence of the linear actuator.

The stability on an equilibrium point for the system can be achieved by feedback controller using direct method and indirect method of Lyapunov. Even though the Lyapunov direct method is an effective way of investigating nonlinear systems and obtaining global results on stability of systems as in, e.g., [40] but instead of looking for a Lyapunov function to be applied directly to the non-linear system, the idea of linearization around a given point is used to achieve stability on some region by using the Lyapunov indirect method. Stability and the stabilization of the system using feedback control law has been analyzed by indirect method and the Jacobian linearization methods in [41]. The results were obtained by using the Lyapunov indirect method to approximate the behaviour of the uncontrolled nonlinear system's trajectory near the critical point using the Jacobian method and designing a state feedback controller for the stabilization of the controlled non-linear system using the difference in response between the set point and actual output values of the system. Furthermore, the Lyapunov-Razumikhin method has been used to determine sufficient conditions for the stabilization of the system using rate of change of a function Rn. Model of Mass spring damper system illustrated the effectiveness of proposed solution by comparing open loop and closed loop results using Matlab.

Inverse optimum control has been proposed as a potential solution to the problem of non-linear system optimal control. The principle of IOC establishes that every stabilizing feedback control optimizes an equivalent performance measure which is determined posterior. Hence unlike the classical approach of optimal control, which depends on the optimization of a priori defined performance measure, IOC approaches the optimal control problem in the converse direction. Inverse optimal control has been introduced in [42] as an alternative strategy for solving the problem of optimal control of non-linear systems where generalized conditions has been proposed for global asymptotic stabilization of closed-loop systems by introducing Control Lyapunov function based Inverse optimal control, combined with State-dependent coefficient matrix for stabilization of second order non-linear systems. The basic purpose of inverse optimal control defined in this study is to compute the stabilizing control action which provides the basis of the system's global asymptotic stabilization and then this state feedback computes performance measures. The restriction on stability by State-dependent Riccati equations control techniques because of their complexity and limitation to only specific second order non-linear systems has been discussed in this note [43, 44]. The proposed global asymptotically stabilization conditions have the advantage of considerable design flexibility as well as application to a wide range of systems.

An alternative solution to the state-space description is operator theoretic viewpoint for the stabilization of non-linear systems by feedback control. The integration of Koopman operator methodology with Lyapunov based model predictive control has been proposed in [45] for stabilization of non-linear systems. For this purpose, the linear Koopman operator theory [46] that could predict upcoming states for non-linear dynamics by observing eigenvalues and eigenfunctions of the operator is discussed. Firstly, the bilinear control model has been developed in Koopman space containing eigenvalues and then the Lyapunov model predictive control method is used to design a closed-loop controller for the stabilization of non-linear dynamical systems. This note introduced a different approach for control method for these systems and presented optimal control solution while considering state and input characteristics for the stability analysis of original nonlinear system. The theoretical solutions were then verified by non-linear Van der Pol oscillator.

## III. Conclusion

The purpose of this review is to briefly discuss some past and recent development in the stabilization of non-linear systems by output feedback control. To control such systems using state/output feedback is a very extensive area for research because of their less complexity in the design procedure as compared to others. The literature review describes various feedback control methods for stabilization and provides an overall summary of the several techniques and their challenges to stabilize and control non-linear systems.